\documentclass[preprint,journal]{vgtc}             

\onlineid{2049}

\preprinttext{To appear in IEEE Transactions on Visualization and Computer Graphics.}

\manuscriptnote{%
  \copyright~2026 IEEE. Personal use of this material is permitted. Permission from
  IEEE must be obtained for all other uses, in any current or future media, including
  reprinting/republishing this material for advertising or promotional purposes,
  creating new collective works, for resale or redistribution to servers or lists,
  or reuse of any copyrighted component of this work in other works.%
}

\vgtccategory{Research}

\title{SocialFiVis: A Visual Analytics Sandbox for \mr{LLM-Grounded} Multi-Agent Simulation in Social Finance}

\author{%
  \authororcid{Yi-Fan Cao}{0000-0002-5892-5052},
  \authororcid{Qing Shi}{0000-0003-2183-6870},
  \authororcid{Liangwei Wang}{0000-0003-3481-3993},
  \authororcid{Leo Yu-Ho Lo}{0000-0002-3660-3765},
  \authororcid{Lin Chen}{0000-0002-2605-749X},
  \authororcid{Yuzi Han}{0000-0002-5952-8593},\\
  \authororcid{Yang Wang}{0000-0002-8903-2388}, and
  \authororcid{Kani Chen}{0000-0003-0117-8065}
}

\authorfooter{%
  \item
    Yi-Fan Cao, Leo Yu-Ho Lo, Yuzi Han, and Kani Chen are with
    The Hong Kong University of Science and Technology (HKUST).
    E-mail: caoyifan@ust.hk, yhload@cse.ust.hk,
    yhanam@connect.ust.hk, and makchen@ust.hk.

  \item
    Qing Shi and Liangwei Wang are with
    The Hong Kong University of Science and Technology (Guangzhou),
    HKUST(GZ). E-mail: \{qshi118,lwang344\}@connect.hkust-gz.edu.cn.

  \item
    Lin Chen is with Northeastern University.
    E-mail: l.chen2@northeastern.edu.

  \item
    Yang Wang is with The University of Hong Kong (HKU).
    E-mail: yang.wang@hku.hk.
}

\abstract{%
The emergence of social finance (SocialFi) transforms online communities into complex socio-economic systems.
\mr{Within these spaces, collective decisions shape a ``digital commons'' characterized by social capital (e.g., community trust) and financial health (e.g., market liquidity).}
Governing such hybrid ecosystems is challenging because real-world interventions are costly and irreversible. While counterfactual simulation is essential for exploring alternative governance strategies, existing approaches fail to capture the non-linear interplay between governance rules, individual behaviors, and emergent economic outcomes.
To systematically unpack this complexity, we operationalize the Institutional Analysis and Development (IAD) framework as our theoretical foundation, synthesizing prior literature with insights from formative expert interviews. Built on this framework, we present \textit{SocialFiVis}, an IAD-embedded visual analytics sandbox.
It introduces a robust model to quantify the dual-track digital commons, coupled with \mr{a two-phase simulation engine.}
\mr{This engine combines LLM-derived personas with a mechanism-guided \textit{Perception--Reasoning--Action} (PRA) runtime to simulate} heterogeneous, context-aware agents \mr{empirically grounded in the retained messaging cohort}.
A hierarchical multi-view interface with interpretable reasoning pathways enables community operators to explore counterfactual policies and trace system-level outcomes back to individual behavioral rationales.
We evaluate \textit{SocialFiVis} through two case studies, a user study, and follow-up interviews.
Results demonstrate that \textit{SocialFiVis} supports fine-grained behavioral attribution and helps explain emergent phenomena such as the structural decoupling of social capital and the resilience of \mr{messaging members} under localized governance shocks.
}

\graphicspath{{figs/}{figures/}{pictures/}{images/}{./}} 

\usepackage{mathptmx} 
\usepackage{amsmath}
\usepackage{xcolor}   
\usepackage{multirow}
\DeclareRobustCommand{\mr}[1]{%
  \texorpdfstring{\textcolor{black}{#1}}{#1}%
}

\usepackage{booktabs}                  

\makeatletter
\g@addto@macro\ps@plain{\let\@evenhead\@oddhead \let\@evenfoot\@oddfoot}
\g@addto@macro\ps@empty{\let\@evenhead\@oddhead \let\@evenfoot\@oddfoot}
\makeatother
\begin{document}

\firstsection{Introduction}
\maketitle

Traditional online communities are gradually evolving from simple social networks into socio-economic systems that incorporate decentralized finance (DeFi) attributes \cite{imani2023socialfi}.
Known as SocialFi, this emerging paradigm \mr{assigns economic value to digital identities and social relationships,} transforming participants into co-owners and governance decision-makers of community assets~\cite{guidi2022socialfi}.
\mr{This dynamic is particularly visible in Non-Fungible Token (NFT) communities~\cite{brahmstaedt2025community}, the empirical setting examined in this work. In these ecosystems, collective decisions shape what the community jointly owns and manages as a ``digital commons''~\cite{li2024governing}.
We characterize this commons through two tightly coupled dimensions: \textit{social capital}, reflected in participation, consensus, and trust~\cite{chiu2006understanding}; and \textit{financial health}, reflected in floor price, liquidity, and the number of holders~\cite{guidi2022socialfi, chen2025decentralized}.}
These dimensions co-evolve; \mr{strong social capital can mitigate} market panic, while \mr{healthy financial conditions sustain participation} incentives~\cite{cao2023nfteller}. 

Governing \mr{this intertwined} digital commons is exceptionally difficult. Communities face rapid \mr{macro-market swings and sentiment shifts~\cite{kapoor2022tweetboost}, leaving \emph{community operators} (e.g., project leads and moderators) to rely on limited experience or intuition when steering governance}~\cite{esposito2025decentralizing}. Furthermore, real-world interventions (e.g., modifying royalty fees or token-gating access) are costly and risky. Testing these governance changes typically requires real budgets, and \mr{poorly designed policies can trigger} irreversible collapse of both social trust and financial liquidity~\cite{li2024governing}. Consequently, a visual sandbox is essential for \mr{operators to explore alternative governance policies} through retrospective counterfactual simulations without consuming actual resources.

However, existing approaches struggle to capture the complex, non-linear interplay between governance rules, individual behaviors, and emergent economic outcomes~\cite{windrum2007empirical}. Traditional \textit{Agent-Based Modeling} (ABM) \mr{provides a principled way to model emergence, but its hand-specified behavioral rules often fall short in reflecting} the nuanced, text-driven social consensus and speculative \mr{behaviors} in online communities.
While Large Language Model (LLM)-based agents can \mr{generate plausible reasoning traces}, current simulations mostly focus on purely conversational networks~\cite{lin2025simspark} or isolated strategic tasks~\cite{chu2025llm}. 
Crucially, they lack \mr{an explicit} institutional structure to \mr{translate} individual motives \mr{into} collective outcomes.
\mr{To bridge this gap, we draw on the Institutional Analysis and Development (IAD) framework~\cite{ostrom2009institutional}, a classic theory explaining how governance rules shape collective behavior and outcomes.
We operationalize this framework as the theoretical backbone of our system,} synthesizing prior literature with insights from a formative expert study.
Yet, realizing this construct within a visual analytics system presents three challenges:

First, defining and quantifying the digital commons is inherently complex.
\mr{Measuring abstract properties like social capital and financial health in SocialFi without standard metrics is non-trivial. Simple data aggregations can easily} mask underlying liquidity crises or be skewed by localized anomalies, necessitating \mr{robust} mathematical modeling~\cite{sanchez2021social}.
Second, achieving ecological validity in \mr{multi-agent simulations} poses severe algorithmic hurdles~\cite{chu2025llm, choi2025proxona}. It requires systematically extracting \mr{data-driven} personas and endowing them with authentic \mr{social and financial} reasoning. Crucially, the simulation architecture must explicitly model how \mr{heterogeneous} individual behaviors adapt to governance rules and aggregate into emergent economic outcomes, thereby capturing their complex, non-linear interplay.
Third, designing intuitive visual interactions for uncertainty-aware governance analysis is demanding. The system must facilitate flexible hierarchical exploration, seamlessly connecting collective policy outcomes with the understandable reasoning pathways of individual agents. Bridging these \mr{macro} consequences and \mr{micro behaviors}, without instilling a false sense of precision, is \mr{vital} for effective decision-making \cite{wang2025framework}.

To address these challenges, we conducted in-depth interviews with domain experts to distill design requirements and define the evaluation metrics for the digital commons. We propose a soft-penalty geometric blend model to robustly quantify social capital and financial health.
\mr{We then developed a two-phase multi-agent simulation engine. Grounded in a seven-dimensional persona codebook, it combines expert-validated, LLM-derived personas with a mechanism-guided runtime to simulate agent behaviors and communication networks.}
Finally, we integrate these models into \textit{SocialFiVis}, an IAD-embedded multi-view visual analytics system for exploring counterfactual outcomes under user-injected policies.
The system comprises an \textit{Event Timeline} for specifying governance interventions, a \textit{Persona View} for examining heterogeneous agent profiles, a \textit{Behavior View} \mr{for revealing behavioral divergence across agent groups}, and a \textit{Communication Network} capturing agent interactions alongside interpretable reasoning pathways to facilitate fine-grained behavior attribution.

\noindent Our primary contributions are summarized as follows:
\begin{itemize}[leftmargin=*, nosep, font=\bfseries]
    \item \textbf{IAD-Driven Socio-Economic Modeling}: \normalfont Based on the IAD framework, we formalize SocialFi communities as a dual-track digital commons and \mr{operationalize its social and financial dimensions through robust quantitative metrics.}
    \item \textbf{\mr{LLM-Grounded} Visual Analytics Sandbox}: \normalfont We develop \textit{SocialFiVis}, a multi-view sandbox \mr{integrating a two-phase simulation engine that combines LLM-derived, expert-validated personas with a mechanism-guided runtime to model how governance rules shape individual behaviors and collective outcomes.}
    \item \textbf{Empirical Insights and Validation}: \normalfont Through case studies and domain expert interviews, we demonstrate the system's efficacy in \mr{revealing} critical governance trade-offs \mr{among messaging members}.
\end{itemize}


\section{Related Work}
\label{sec:related_work}

\mr{We situate this research at the intersection of three areas: the governance of SocialFi digital commons, agent-based social simulation}, and visualization for counterfactual reasoning.

\subsection{SocialFi and Digital Commons Governance}
SocialFi represents the convergence of social networks and tokenized economies, where digital assets (e.g., NFTs) \mr{carry both economic value and governance rights}, transforming users into co-owners and decision-makers of community resources~\cite{guidi2022socialfi, ricci2024awesome}.
In these ecosystems, the shared resources managed by members constitute a ``digital commons'' featuring two tightly coupled dimensions: \textit{social capital}, \mr{commonly characterized by participation, consensus, and trust~\cite{chiu2006understanding, jeong2021measure};} and \textit{financial health}, \mr{typically comprising asset floor price, liquidity, and holder count~\cite{chen2025decentralized} (both quantified in Sec.~\ref{sec:model}).}
Empirical evidence shows that sustained engagement in such communities relies heavily on social identity formation rather than purely financial speculation, necessitating governance structures that balance both objectives~\cite{brahmstaedt2025community}. To analyze this dual-track commons, we draw upon Elinor Ostrom's Institutional Analysis and Development (IAD) framework, \mr{which explains how governance rules, actors, and shared resources interact to shape collective outcomes}~\cite{ostrom2009institutional, mcginnis2019connecting}.
While the IAD framework has been widely applied to online creation communities~\cite{morell2014governance, schlager2018iad}, existing quantification efforts in SocialFi predominantly rely on ad-hoc, isolated indicators (e.g., tracking price or sentiment independently). These approaches fail to capture the systemic vulnerability where the collapse of one dimension catastrophically impacts the overall ecosystem.
\mr{Building on the IAD framework, we propose a soft-penalty geometric blend model that robustly quantifies the dual-track digital commons.}

\subsection{Agent-Based Modeling and \mr{LLM-Based Simulation}}
ABM provides a principled bottom-up framework for studying how individual behavioral rules, under institutional constraints, aggregate into macro-level social dynamics~\cite{bonabeau2002agent, epstein1999agent, macal2010tutorial}.
\mr{LLM-based} agents have recently extended classical ABM by replacing hand-coded heuristics with language-grounded reasoning, enabling emergent sociolinguistic behaviors in simulated communities~\cite{park2023generative, gao2024llm, gurcan2024llm, chen2025ai}. \mr{While they have shown promise as behavioral proxies in socio-economic settings}~\cite{filippas2024large, ziems2024llm}, \mr{validating} simulation fidelity against real-world trajectories remains challenging due to the stochastic nature of LLM outputs~\cite{windrum2007empirical, fagiolo2007critical, guerini2017method, barde2017empirical, larooij2025validation}. 
\mr{One promising direction is to improve ecological validity by initializing agents with empirically grounded personas derived from community interaction data~\cite{choi2025proxona}.
While these approaches better reflect the diversity of real community members, they continue to model agent interactions without explicit institutional structures, focusing solely on conversational exchanges~\cite{tang2025gensim,zhang2025socioverse} or isolated market behaviors~\cite{chu2025llm}.
Consequently, how institutional governance shapes the transition from individual decisions to collective outcomes remains largely unexplored~\cite{taillandier2025integrating}.}
\mr{We address this gap with a two-phase architecture: LLM-derived personas establish heterogeneous agent motives, while a mechanism-guided runtime operationalizes IAD governance rules to systematically bridge this micro-macro divide.}

\subsection{Visualization for Counterfactual Reasoning}
Visualization research for financial technology has matured across distinct analytical tasks. Existing systems predominantly address transaction network analysis~\cite{tovanich2019visualization, zhong2020silkviser, cao2025nftracer}, market metric monitoring~\cite{ni2023using}, and fraud detection~\cite{zhou2023visual, wen2024ponzilens+j}. However, these tools typically lack a governance context, treating transactions as isolated events rather than outcomes of collective decision-making. Social visualization research has explored the relationship between online engagement and market dynamics, including sentiment contagion mapping~\cite{li2025causalmap}, but often overlooks the institutional rules shaping participant behavior.
Closer to our goal, visual causal inference has established counterfactual reasoning as a means of examining how interventions shape observed outcomes~\cite{wang2024empirical, borland2024using, sohns2023decision, wang2025framework, gajcin2024redefining, kaul2021improving}. Despite these advances, few systems support policy backtesting with traceability from macro outcomes to micro behavioral rationales~\cite{ge2024vframer}.
Existing agent-based social simulation platforms likewise tend to emphasize aggregate outcomes over drill-down inspection of individual agents' decision pathways~\cite{tang2025gensim,zhang2025socioverse}. Such aggregate presentation also carries a documented risk of false precision, as deterministic visual encodings can project unwarranted certainty over stochastic outcomes~\cite{atrey2019exploratory}.
We address these limitations through \textit{SocialFiVis}, which integrates IAD-based policy configuration, multi-agent simulation, and hierarchical visual analytics with interpretable reasoning traces for uncertainty-aware counterfactual reasoning.

\begin{figure*}[tb]
\centering
\includegraphics[width=\textwidth,
  alt={Overview of the IAD-embedded SocialFiVis pipeline, including data processing, persona extraction, simulation, and visual analysis.}]{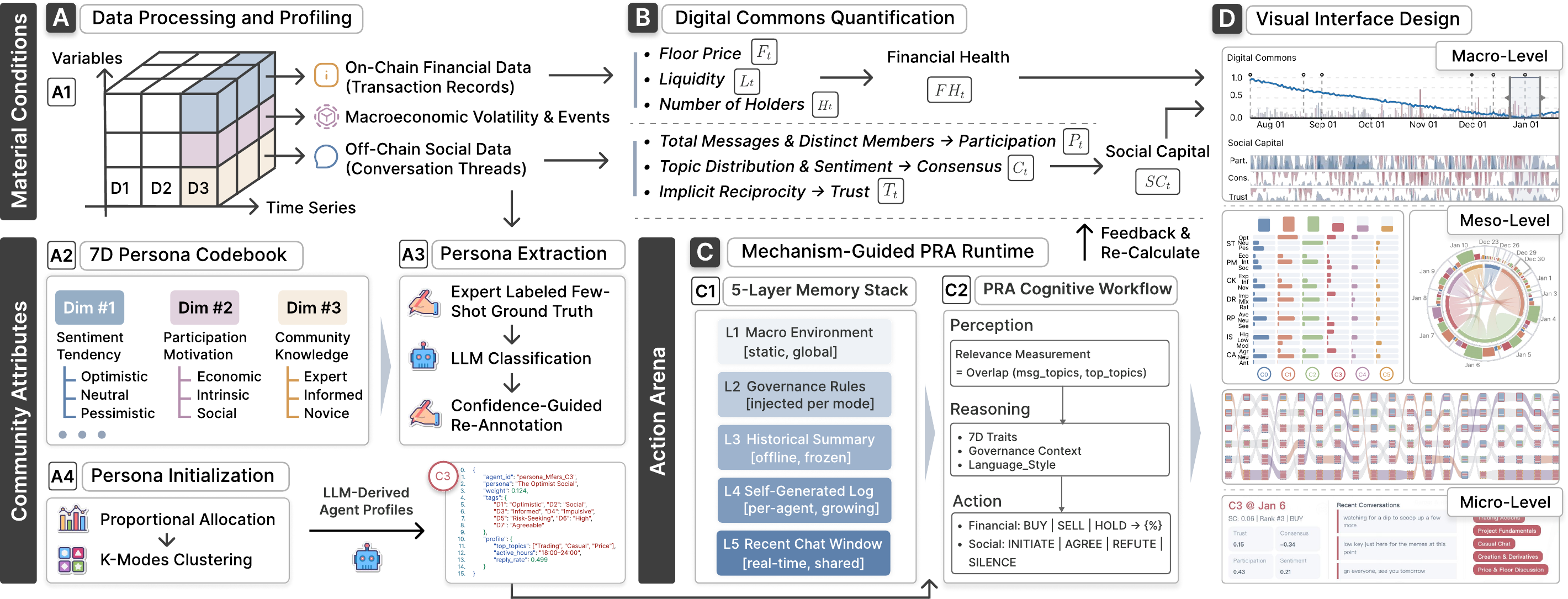}
\caption{IAD-embedded system overview. (1) Data Processing and Quantification (A, B) establish baseline \textit{Material Conditions} and extract \textit{Community Attributes}. (2) An \textit{Action Arena} (C) simulates agent behaviors under counterfactual policies. (3) The Visual Interface (D) presents \textit{Evaluative Outcomes}. A closed feedback loop (C $\rightarrow$ B) aggregates simulated micro-actions to iteratively update the \textit{Material Conditions} tick-by-tick.}
\vspace{-2mm}
\label{fig:overview}
\end{figure*}

\section{Formative Study}

To ground the system \mr{in practice}, we adopted a participatory design approach \mr{centered on \emph{community operators} (project leads and moderators) who diagnose community conditions and compare governance interventions. Analysts, traders, and community members are secondary users who interpret or audit these decisions. We collaborated with three practitioners who have hands-on SocialFi community governance experience:} $E_1$ (a former project lead), $E_2$ (COO of a crypto AI infrastructure project), and $E_3$ (a core community contributor and institutional liaison). A computational social science and ABM researcher ($E_4$) \mr{served as a methodological co-design advisor. Interviews with $E_1$--$E_3$ surfaced} the governance workflows and operational bottlenecks.

\subsection{Design Goals}

\mr{Synthesizing the needs of our practitioners ($E_1$--$E_3$) with methodological input from our simulation advisor ($E_4$),} we distilled three high-level design goals (DGs) to guide our system development.

\textbf{DG1: Quantify the Coupling Between Social and Financial Dynamics.} All \mr{practitioners} emphasized that social behaviors and financial metrics are deeply intertwined yet \mr{are tracked in isolation}. $E_1$ noted that these data streams are ``\textit{completely siloed,}'' lacking a unified tool to examine their underlying coupling. $E_3$ echoed this, describing community dynamics and financial outcomes as ``\textit{mutually influencing each other.}'' The system must integrate and visualize the co-evolution of social capital and financial health within a unified analytical framework.

\textbf{DG2: Enable Counterfactual Policy Evaluation With Retrospective Grounding.} \mr{The practitioners} expressed a critical need to move from post-hoc analysis toward simulation-based strategy assessment.
$E_2$ described her current workflow as ``\textit{entirely reactive},'' \mr{while $E_1$ sought to ``\textit{back-test strategies}'' rather than rely on intuition}.
From a modeling perspective, $E_4$ \mr{emphasized that assessing alternative interventions requires a controlled, historically grounded counterfactual simulation.
The system must therefore enable operators to compare governance policies in a risk-free sandbox.}

\textbf{DG3: Support Heterogeneous User Profiling and Behavioral Attribution.} \mr{The practitioners stressed that modeling} the diverse behavioral patterns within SocialFi networks is critical for understanding market shifts. Rather than treating the community homogeneously, $E_2$ and $E_3$ \mr{highlighted the need to examine} how specific cohorts react differently to market events and governance changes. To operationalize this for simulation, $E_4$ identified the central technical challenge as ``\textit{capturing the heterogeneity among community members... to faithfully reproduce macro-level dynamics.}'' Thus, the system must derive distinct persona types from empirical data and trace how macro-level outcomes emerge from their differentiated behaviors.

\subsection{\mr{The IAD Framework as an Analytical Lens}}

\mr{To achieve the goal of a unified framework that disentangles} social and financial dynamics, we adopted Elinor Ostrom's IAD framework~\cite{ostrom2009institutional}.
\mr{We selected this theoretical lens in close consultation with our practitioners because it directly addresses their analytical needs.} $E_1$ noted that Ostrom's focus on the decentralized governance of common-pool resources naturally aligns with the organizational logic of SocialFi communities.
Furthermore, $E_2$ and $E_3$ \mr{valued the framework's analytical cycle~\cite{mcginnis2019connecting}, which supports quantitative attribution by} mapping exogenous variables to action situations and evaluating resulting outcomes.

$E_4$ \mr{further emphasized that the strength of the IAD framework lies in its systematic translation of institutional concepts into computational components. This principled mapping allows governance rules, community attributes, and environmental conditions to be operationalized consistently across the two-phase multi-agent simulation.}

\subsection{Analytical Tasks}

Building on the design goals, \mr{we refined six analytical tasks (\textbf{T1}--\textbf{T6}) through iterative discussions with $E_1$--$E_3$. Following the IAD framework, we organize these tasks across three nested analytical scales (macro, meso, and micro).
These scales mirror how operators naturally reason about their communities: from diagnosing aggregate outcomes, to identifying cohort-specific reactions, and finally tracing individual motives to overcome the anonymity of on-chain data.}

\noindent\textbf{Macro-Level Community Dynamics.} \mr{Operators} seek to understand how governance configurations dictate aggregate outcomes.
\begin{itemize}[leftmargin=*, noitemsep, topsep=2pt]
    \item \textbf{T1. Temporal Metric Exploration (DG1).} Quantify and visually track the dual-track digital commons (social capital and financial health). The system must support interactive exploration of temporal patterns to ground reasoning \mr{about commons} evolution.
    \item \textbf{T2. Cross-Metric Policy Impacts (DG2).} Support comparative sandbox analysis to examine the interplay between metrics under varying rules-in-use. This enables users to test alternative governance configurations and \mr{observe their systemic effects}.
\end{itemize}

\noindent\textbf{Meso-Level Persona Patterns.} \mr{Operators} investigate how large-scale, heterogeneous participation drives these macro-level shifts.
\begin{itemize}[leftmargin=*, noitemsep, topsep=2pt]
    \item \textbf{T3. Data-Driven Persona Derivation (DG3).} Derive representative persona types from empirical behavioral and investment patterns, surfacing influential cohorts and their characteristic behaviors to support heterogeneity-aware analysis.
    \item \textbf{T4. Behavior-Outcome Backtesting (DG2, DG3).} Enable retrospective analysis to map specific persona actions to macro-level shifts,
    \mr{uncovering behavioral drivers underlying fluctuations in the digital commons}.
\end{itemize}

\noindent\textbf{Micro-Level Cognitive Pathways.} \mr{Operators} aim to interpret the latent decision-making rationales that shape observed behaviors.
\begin{itemize}[leftmargin=*, noitemsep, topsep=2pt]
    \item \textbf{T5. Policy-Behavior Differentiation (DG2, DG3).} Reveal how shifts in rules-in-use trigger heterogeneous responses across distinct persona types, exposing specific behavioral nuances that are often obscured in aggregate metrics.
    \item \textbf{T6. Decision Pathway Interpretation (DG1, DG3).} Unpack the \mr{persona-conditioned} reasoning pathways of agent types to offer an explanatory context for why specific behaviors emerge, bridging actions with underlying rationales.
\end{itemize}

\section{System Overview}

Guided by the analytical tasks, \mr{we instantiate the IAD framework as} an interactive visual analytics sandbox consisting of three tightly coupled modules (Fig.~\ref{fig:overview}).
(1) The data processing and quantification module establishes the system's empirical baseline. It parses heterogeneous on-chain and off-chain data to construct the \textit{Material Conditions}, which quantify the community's social capital ($SC_t$) and financial health ($FH_t$). Meanwhile, it distills the \textit{Attributes of the Community} by clustering users into representative archetypes using a seven-dimensional persona codebook. 
(2) The \mr{two-phase} multi-agent simulation engine serves as the \textit{Action Arena}. 
\mr{It pairs LLM-derived personas with a mechanism-guided runtime executing a} \textit{Perception--Reasoning--Action} (PRA) pipeline.
\mr{Conditioned on} user-injected counterfactual governance policies (\textit{Rules-in-Use}), \mr{agents produce} heterogeneous social and financial interactions (\textit{Action Situations}).
(3) The visual analytics dashboard presents the \textit{Evaluative Outcomes} through coordinated views that support multi-level (macro--meso--micro) policy exploration. 
Crucially, within any selected time window, these three modules form a closed-loop co-evolutionary system. Simulated micro-actions are mathematically aggregated to perturb the baseline $SC_t$ and $FH_t$ tick-by-tick, continuously updating the \textit{Material Conditions} for the internal ongoing simulation. This architecture empowers target users to dynamically evaluate the cascading impacts of governance interventions.

\section{Data Analysis}

This section details the collection of heterogeneous data and the mathematical formulation of community metrics.

\subsection{Data Collection and Preprocessing}
\label{sec: data}

We constructed a heterogeneous dataset capturing the digital footprints of SocialFi communities across a ten-month period (Jul. 2022--Apr. 2023), integrating three core contexts (Fig.~\ref{fig:overview}A1):

\textbf{Off-Chain Social Data}: We collected chat logs from the official communities of two representative SocialFi projects during their respective active phases. \textit{Mfers}~\cite{mfers} was selected as a Western blue-chip project demonstrating sustained market presence, while \textit{Mimic Shhans}~\cite{mimicshhans} represents a prominent Asian project captured during its accelerated growth phase. 

\textbf{On-Chain Transaction Data}: Utilizing the NFTGo and OpenSea APIs, we extracted daily financial metrics for both projects, including \textit{floor price} ($F_t$), \textit{liquidity} ($L_t$), and the \textit{number of holders} ($H_t$), where $t$ denotes the daily time step. All values were normalized to a $[0, 1]$ scale.

\textbf{Macroeconomic Environment}: We compiled a chronological dataset of major events impacting the broader SocialFi market (e.g., Federal Reserve rate hikes, the FTX collapse) sourced from authoritative platforms (CoinDesk, Bloomberg). This serves as the global context layer for our simulation.


\subsection{Mathematical Modeling of Community Metrics}
\label{sec:model}

To translate unstructured behaviors into quantifiable time-series, we formalize social capital ($SC_t$) and financial health ($FH_t$) through a tri-dimensional vector space (Fig.~\ref{fig:overview}B).
This balanced design prevents index skewing by localized anomalies, providing a robust, multidimensional foundation for subsequent visual encoding.

\subsubsection{Social Capital ($SC_t$) Modeling}
Serving as a robust proxy for intangible community cohesion, $SC_t$ quantifies the structural, cognitive, and relational resilience that sustains a project's long-term viability beyond mere financial speculation.
We strictly map daily $SC_t$ to a $[-1, 1]$ interval. $SC_t$ comprises three metrics:

\textit{Participation ($P_t$)} captures structural engagement~\cite{jones2011metrics}. Relying solely on message frequency invites spamming by a few hyperactive users~\cite{papakyriakopoulos2020hyperactive}. Thus, we incorporate $N_t^{unique}$ (the number of \mr{distinct daily senders}) alongside total messages ($N_t^{msg}$). \mr{Here $\mu_{msg}$ is the community's historical mean daily message volume, a static baseline for activity fluctuations.} \mr{A $\tanh$ caps activity spikes; the two ratios carry equal weight (high scores need both intensity and breadth), and the $0.5$ offset centers a typical day near zero}:
\begin{equation}
P_t = \tanh \bigg( \ln \Big( 1 + \frac{N_t^{msg}}{\mu_{msg}} \cdot \frac{N_t^{unique}}{N_{members}} \Big) - 0.5 \bigg)
\end{equation}

\textit{Consensus ($C_t$)} deconstructs cognitive consensus into topical and emotional dimensions~\cite{blei2003latent, griffiths2004finding}. For the topical dimension, we apply LDA to obtain a daily distribution over $n$ topics, $p_1, \dots, p_n$, and compute its Shannon entropy $\mathcal{H}_t = -\sum_{i=1}^n p_i \ln(p_i)$. A lower $\mathcal{H}_t$ denotes discourse concentrated on a few topics (high consensus), whereas entropy approaching the maximum $\ln(n)$ reflects fragmented attention. For the emotional dimension, we compute the standard deviation ($\sigma_t$) of daily sentiment scores from a domain-adapted, SocialFi-tuned classifier; a low $\sigma_t$ implies strong emotional alignment even amid widespread panic.
\mr{We weight the two dimensions equally, so that high consensus requires both focused discourse and emotional alignment}:
\begin{equation}
C_t = 0.5 \cdot \left(1 - 2 \frac{\mathcal{H}_t}{\ln(n)}\right) + 0.5 \cdot \left(1 - 2\sigma_t\right)
\end{equation}

\textit{Trust ($T_t$)} adapts to platforms lacking explicit reply structures~\cite{al2012group}.
\mr{We define \textit{implicit reciprocity} ($IR_t$) as the density of reciprocated directed edges among members, established by sequential, topically related posts within a five-minute sliding window and normalized to $[-1,1]$.}
$T_t$ is formalized as an equal-weighted blend of $IR_t$ and the daily mean sentiment polarity ($\mu_t$).
\mr{This balancing mechanism serves as a safeguard: high trust scores require both constructive engagement and non-negative sentiment, filtering out ``flame wars'' where turn-taking may be frequent} but sentiment polarity is negative:
\begin{equation}
T_t = 0.5 \cdot IR_t + 0.5 \cdot \mu_t
\end{equation}

To synthesize the overall social capital ($SC_t$), we first normalize $P_t, C_t$, and $T_t$ to $[0, 1]$. We then apply a \textit{soft-penalty geometric blend} to formulate the intermediate index $SC'_t$.
\mr{By the \textit{arithmetic mean--geometric mean} (AM--GM) inequality, a purely geometric aggregate can over-penalize routine sub-metric variation, shifting $SC_t$ downward relative to arithmetic aggregation and, in our data, toward pessimistic values. We therefore set $\alpha=0.5$ as a symmetric blend, tempering this bias while retaining geometric weak-link sensitivity to sudden crises:}
\begin{equation}
SC'_t = \alpha \cdot \left( \frac{P'_t + C'_t + T'_t}{3} \right) + (1 - \alpha) \cdot \sqrt[3]{P'_t \cdot C'_t \cdot T'_t}
\end{equation}
where $X'_t = (X_t + 1) / 2$ for $X \in \{P, C, T\}$. The result is subsequently mapped back to the diverging $[-1, 1]$ visual space:
\begin{equation}
SC_t = 2 \cdot SC'_t - 1
\end{equation}

\subsubsection{Financial Health ($FH_t$) Modeling}
Derived from on-chain transaction data (Sec.~\ref{sec: data}), $FH_t$ evaluates market vitality on a strict $[0, 1]$ scale. Informed by domain experts, we synthesize three orthogonal metrics, \textit{floor price} ($F_t$), \textit{liquidity} ($L_t$), and \textit{holder count} ($H_t$).
\mr{As these metrics are individually normalized and no domain rationale privileges one, we combine them with an \emph{unweighted} geometric mean. The geometric form enforces a ``shortboard penalty'': a collapse in any component triggers a non-linear plunge, mitigating the ``paper-wealth trap'' in which an artificially high floor price masks a liquidity crisis, and keeping isolated anomalies visually salient}:
\begin{equation}
FH_t = \sqrt[3]{F_t \cdot H_t \cdot L_t}
\end{equation}

\subsubsection{\mr{Sensitivity and Robustness Analysis}}
\mr{We assess the effect of these parameter choices through sensitivity analyses. Sweeping $\alpha \in [0,1]$ preserves the temporal ordering of $SC_t$ ($\rho \geq 0.991$), showing that the observed $SC_t$ trends are not sensitive to this midpoint setting. Reweighting the three financial inputs preserves the $FH_t$ ranking ($\rho \geq 0.988$).
For \mr{$P_t$,} $C_t$ and $T_t$, moderate alternatives around the equal split ($w\in[0.25,0.75]$) retain substantial downstream $SC_t$ rank agreement ($\rho\geq0.913$)\mr{; the $P_t$ centering offset ($c\in[0.3,0.7]$) is likewise immaterial ($\rho\geq0.999$)}.
The \textit{Supplement} verifies that these perturbations preserve the key case-level $SC_t$--$FH_t$ interpretations.}

\section{\mr{Two-Phase LLM-Grounded} Multi-Agent Simulation}

\mr{Our sandbox enables counterfactual simulation through a two-phase architecture: an LLM derives expert-validated personas, and a mechanism-guided PRA runtime executes agent interactions without per-tick LLM calls. This separation grounds agent behavior in real community semantics while keeping the runtime controllable, auditable, and fast enough for interactive analysis.}

\subsection{\mr{Phase~I:} Contextual Persona Extraction}
\label{sec: persona extra}

To simulate counterfactual governance scenarios, we must instantiate agents with empirically grounded behaviors~\cite{choi2025proxona}. To prevent cross-contamination, we extracted personas from the two communities independently. Informed by domain experts and sociological literature, we defined a closed-set codebook across seven 3-class dimensions: \textit{sentiment tendency, participation motivation, community knowledge, decision rationality, risk preference, influence seeking,} and \textit{communication agreeableness} \mr{(see Fig.~\ref{fig:expert_validation} in Appendix)}.

To ensure coding reliability, we deployed a confidence-guided human-in-the-loop (HITL)~\cite{mosqueira2023human} pipeline. After experts annotated a few-shot ground truth, a long-context LLM (Moonshot API) labeled the remaining users.
\mr{We excluded silent and near-silent accounts ($<15$ messages), whose sparse text provides insufficient evidence for reliable persona inference. The retained cohort comprises about half of all accounts (see Table~\ref{tab:filtering_stats} in Appendix) while contributing over 95\% of all messages in both communities.}
The LLM outputs were triaged by confidence: high-confidence samples were auto-retained, while uncertain cases (confidence $< 0.60$) underwent expert re-annotation. This data-centric iterative refinement maintained high annotation consistency without requiring model retraining.

Finally, we applied \textit{K}-Modes~\cite{huang1998extensions, huang1999fuzzy} clustering to the labeled features. Based on multi-criteria evaluations (inertia, silhouette score, and ARI), we identified $K = 6$ for \textit{Mfers} and $K = 7$ for \textit{Mimic Shhans} as optimal balances of compactness and interpretability.
Robustness checks confirmed that the dominant archetypes remain stable across random seeds.
These structural archetypes were then enriched with dynamic traits extracted from chat logs, such as \textit{top topics, active routine, reply rate,} and \textit{language style}. Together, they form the foundational templates for agent instantiation. This synthesis of structural archetypes and nuanced linguistic styles \mr{helps reduce} generic ``AI tone,'' \mr{yielding authentic, domain-specific language for the \textit{Phase~II} runtime.}

\begin{figure*}[t]
\centering
\includegraphics[width=\textwidth,
  alt={SocialFiVis interface showing the control panel, persona view, event timeline, behavior view, communication network, and reasoning pathways.}]{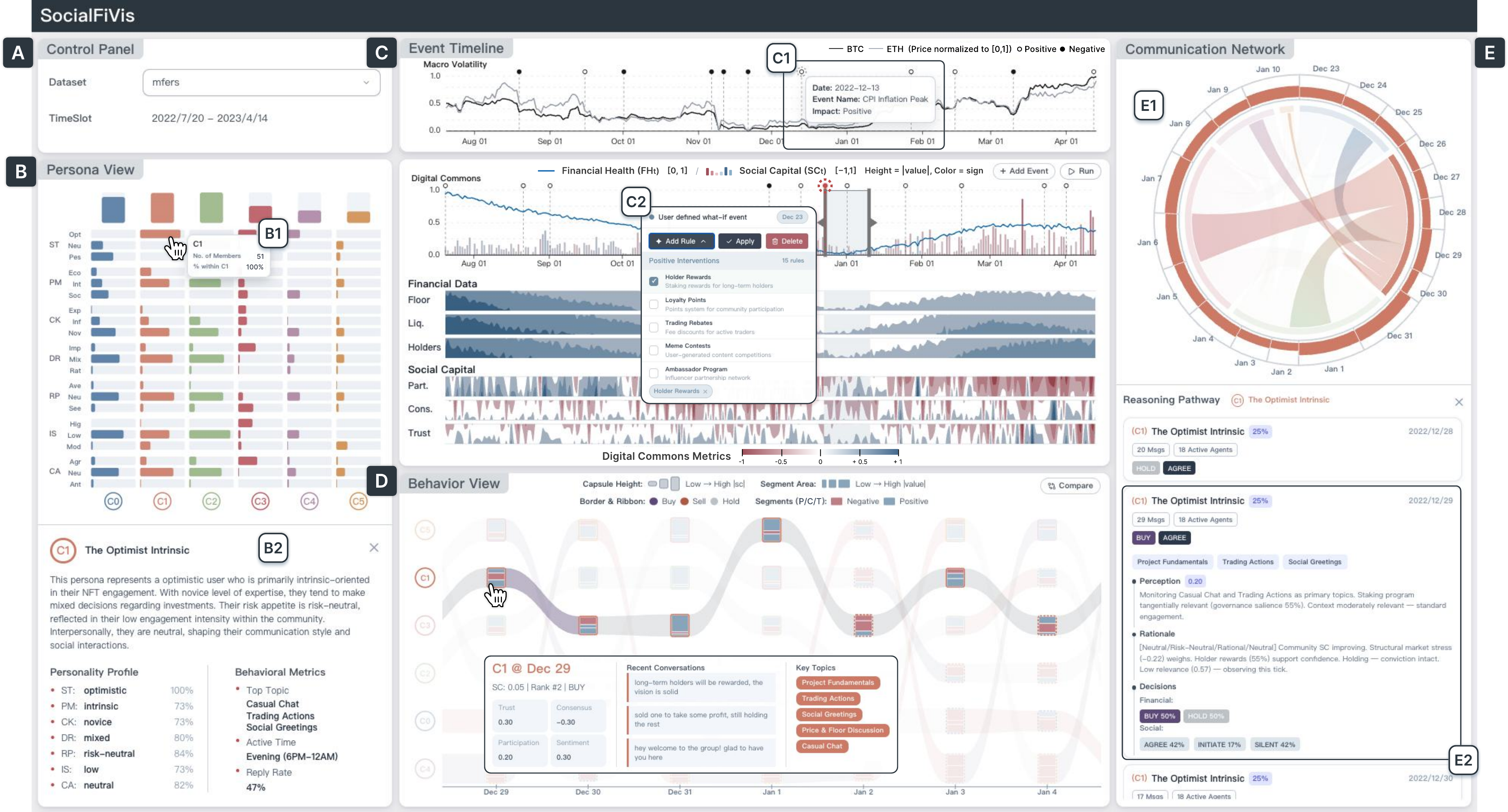}
\caption{The \textit{SocialFiVis} interface. \mr{(A) \textit{Control Panel} for case selection; (B) \textit{Persona View} of persona--trait compositions, with (B1) population details on hover and (B2) a \textit{Persona Card} of profile and behavioral metrics; (C) \textit{Event Timeline} of (C1) macro crypto trends and milestones over community metrics, with (C2) user-injected interventions; (D) \textit{Behavior View} tracking persona trajectories; and (E) \textit{Communication Network} and \textit{Reasoning Pathway} pairing (E1) the communication topology with (E2) per-agent reasoning.}}
\vspace{-2mm}
\label{fig:interface}
\end{figure*}

\subsection{\mr{Phase~II:} Simulation Engine and Cognitive Workflow}

Building upon interactive social media simulation frameworks~\cite{lin2025simspark}, we design a structured \textit{Perception--Reasoning--Action} (PRA) pipeline to govern agent behavior over time (Fig.~\ref{fig:overview}C).
The simulation progresses in discrete time steps (ticks), with each step representing a daily window to capture fine-grained social dynamics.

\begin{table}[tb]
\centering
\caption{Simulation validation results: Mode$_0$ vs. ground truth.}
\label{tab:validation}
\setlength{\tabcolsep}{3.5pt} 
\resizebox{\columnwidth}{!}{
\begin{tabular}{@{}l ccc ccc cc c@{}}
\toprule
 & \multicolumn{3}{c}{\textbf{$SC_t$}} & \textbf{$P_t$} & \textbf{$C_t$} & \textbf{$T_t$} & \multicolumn{2}{c}{\textbf{$FH_t$}} & \textbf{Overall} \\
\cmidrule(lr){2-4} \cmidrule(lr){5-7} \cmidrule(lr){8-9}
\textbf{Community} & $\rho$ & DTW & JSD & $\rho$ & $\rho$ & $\rho$ & $\rho$ & JSD & \textbf{Fidelity} \\
\midrule
\textit{Mfers} & 0.877 & 0.109 & 0.389 & 0.910 & 0.758 & 0.421 & 1.000 & 0.000 & 0.746 \\
\textit{Mimic Shhans} & 0.865 & 0.095 & 0.530 & 0.917 & 0.902 & 0.739 & 1.000 & 0.000 & 0.759 \\
\bottomrule
\multicolumn{10}{@{}l@{}}{\rule{0pt}{2.5ex}\scriptsize $^{*}$ All $p < 0.01$ (except $T_t$ in \textit{Mfers}: $p = 0.073$). $FH_t$ matches ground truth by design in Mode$_0$.}
\end{tabular}%
}
\end{table}

\subsubsection{Simulation Engine Pipeline}

The composition of the simulated population is determined by two constraints. To preserve ecological validity, agents are allocated in proportion to the empirical distribution of \textit{K}-Modes archetype clusters. This choice follows a core principle in agent-based modeling: emergent phenomena, such as influence propagation and consensus formation, are shaped by the relative prevalence of agent types; deviating from this distribution (e.g., via uniform allocation) would systematically bias collective outcomes~\cite{epstein1996growing, bonabeau2002agent, macal2014introductory}.
At the same time, to ensure statistically stable estimation of archetype-level behavioral patterns, we enforce a minimum of $N_p \geq 5$ agents per archetype~\cite{cochran1952chi2, agresti2013categorical}.
Together, these constraints establish a principled lower bound on the total population.

Within this temporal structure, interactions unfold asynchronously rather than through global synchronization.
\mr{To reflect this property, we introduce a \textit{coordinator} that selects the next active agent from the evolving conversational context.}
The coordinator does not generate content; it only regulates turn-taking, thereby preserving realistic communication dynamics without interfering with agent behavior.

Each agent operates with a structured memory comprising five layers: (1) macro environment, (2) user-injected governance rules, (3) historical summaries, (4) self-generated past statements, and (5) the recent chat window. These layers jointly support consistent behavior under long-context interactions. 
Conditioned on this memory, agents execute the PRA pipeline in a tightly coupled manner: during \textit{Perception}, incoming information is filtered according to archetype-specific topical relevance;
\mr{during \textit{Reasoning}, agents weigh their seven-dimensional traits against persona-specific behavioral thresholds;} and during \textit{Action}, they \mr{select} financial decisions (\textit{buy}, \textit{sell}, \textit{hold}) and social interactions (\textit{initiate}, \textit{agree}, \textit{refute}, \textit{silence}), \mr{with posts drawn from context-conditioned, LLM-distilled message templates}.
Relational social actions (e.g., \textit{agree}, \textit{refute}) are subsequently mapped to directed interactions, yielding a daily-resolved network that directly drives the frontend visualization and aligns with downstream financial metrics.

These simulated actions are then fed back into our mathematical models (Sec.~\ref{sec:model}), continuously recalculating the dynamic shifts in $SC_t$ and $FH_t$ to complete the visual analytics loop. Specifically, aggregated financial actions (\textit{buy}/\textit{sell}) perturb the baseline on-chain floor price ($F_t$) and liquidity ($L_t$) metrics. Simultaneously, the newly generated semantic edges alter the structural implicit reciprocity ($IR_t$), while the simulated agent statements shift the collective sentiment variance ($\sigma_t$). This closed-loop mechanism \mr{enables} users to visually trace how micro-level interventions cascade through individual cognition to drive macro-level shifts in community vitality.

\subsubsection{Simulation Evaluation}

We iteratively refined the simulation to improve ecological validity, correcting systemic issues including zero sell-rate bias, absence of bear-market baselines, and missing counterfactual reference modes. The final version introduces GT-anchored environmental calibration, a standard ABM practice \cite{fagiolo2007critical, windrum2007empirical} wherein empirically observed activity levels, sentiment, and topic concentration serve as exogenous environmental inputs. To validate fidelity, we compare Mode$_0$ (no governance) against empirical ground truth using three complementary metrics: Spearman's $\rho$ for trend fidelity \cite{windrum2007empirical}, normalized Dynamic Time Warping (DTW) for shape similarity \cite{berndt1994using}, and Jensen--Shannon Divergence (JSD) for distributional fidelity \cite{endres2003new}. Results confirm strong trend fidelity for core metrics across both communities (see details in~\cref{tab:validation}), with overall composite fidelity scores of $0.746$ and $0.759$ respectively.
\mr{Trust ($T_t$) is the only metric not reaching significance in \textit{Mfers} ($\rho=0.421$, $p=0.073$), which may reflect community-specific language patterns whose subtle trust cues are smoothed by LLM-distilled message templates~\cite{park2023generative, gao2024llm}.
This limits trust-specific interpretation for \textit{Mfers}, although the aggregate $SC_t$ trajectory remains well aligned with ground truth ($\rho=0.877$), supported by stronger recovery of $P_t$ and $C_t$.}
$FH_t$ achieves perfect calibration in Mode$_0$ by design, as it uses unmodified real-world financial data, establishing a principled baseline against which governance interventions are evaluated.

\begin{figure*}[tb]
\centering
\includegraphics[width=\textwidth,
  alt={Behavior View workflow comparing ground-truth community dynamics with counterfactual persona trajectories after an injected event.}]{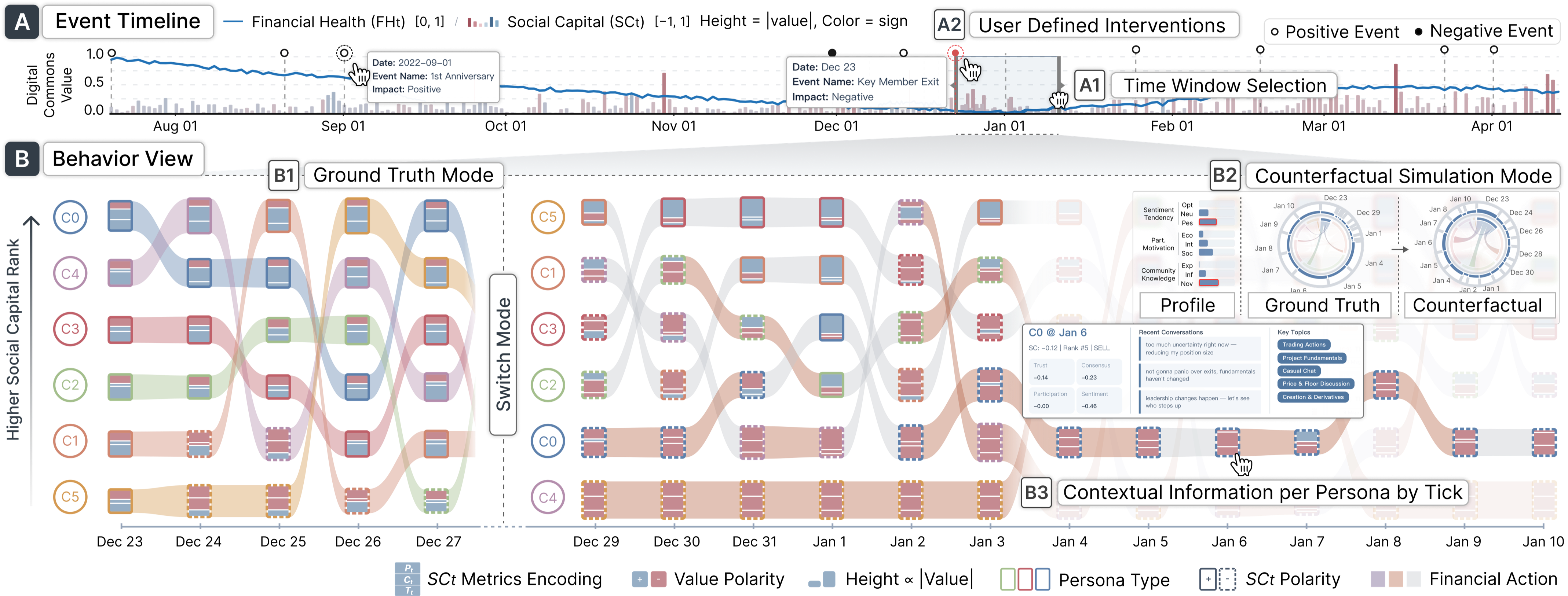}
\caption{Behavior View workflow. (A1) Time window selection loads ground truth community dynamics (B1), while (A2) injecting a negative event (e.g., ``Key Member Exit'') generates counterfactual socio-financial behaviors (B2). Persona highlighting (B3) enables micro-analysis: reacting to the shock, pessimistic C0 agents display heightened, negative social engagement coupled with reactive ``SELL'' strategies (orange ribbons).}
\vspace{-2mm}
\label{fig:behaviorview}
\end{figure*}

\section{Visual Design}

This section presents the visual design of \textit{SocialFiVis}, which consists of four coordinated views designed to support the hierarchical workflow.

\subsection{Control Panel and Persona View}

The \textit{Control Panel} serves as the entry point for scenario selection and data instantiation (Fig.~\ref{fig:interface}A). Selecting a preprocessed case determines the temporal scope and the community, triggering the persona construction whose results are visualized in the \textit{Persona View} (Fig.~\ref{fig:interface}B).

The \textit{Persona View} is a matrix-based visualization that characterizes the extracted persona clusters and their multi-dimensional trait compositions (\textbf{T3}). We employ a structured grid layout. Each column represents a distinct persona, assigned a consistent categorical color to preserve identity across the system. The rows correspond to the seven personality dimensions (as defined in Sec.~\ref{sec: persona extra}). Within each intersecting cell, a group of horizontal bars encodes the probability distribution of the three categorical states for that specific dimension. Additionally, the solid-colored block at the top of each column indicates the persona's relative population size, which directly determines the agent-proportion initialization in the downstream simulation.
\mr{The matrix extends horizontally with the number of personas; once it exceeds the palette's distinguishable hues, personas are grouped into higher-level archetypes.}

Interaction enables progressive disclosure of agent details. Hovering over a horizontal bar triggers an overlay showing the absolute member count and its relative percentage within the persona (Fig.~\ref{fig:interface}B1). Selecting a persona column opens a \textit{Persona Card}, which provides a textual summary of the overarching archetype~(Fig.~\ref{fig:interface}B2). The card further decomposes the abstract configuration into a detailed \textit{Personality Profile}, listing the dominant state and its exact percentage per dimension, alongside grounded \textit{Behavioral Metrics}, such as primary discussion topics, peak active hours, and reply rates. This design bridges multi-dimensional trait distributions with interpretable behavioral descriptors.

\subsection{Event Timeline}

The \textit{Event Timeline} supports the temporal exploration of the digital commons within a macroeconomic context (\textbf{T1}) and serves as the primary sandbox interface for comparative policy evaluation (\textbf{T2}). It helps users pinpoint critical historical milestones and transition fluidly from observational analysis to counterfactual backtesting.

The visual design adopts a multi-tier layout that structurally decouples macroeconomic context from localized community dynamics (Fig.~\ref{fig:interface}C). The top tier tracks global cryptocurrency trends using line charts, with historical milestones embedded as circular glyphs: hollow circles for positive events and solid black dots for negative ones (Fig.~\ref{fig:interface}C1). The middle tier depicts the community's overarching trajectory via a primary line chart paired with diverging bars. The bottom tier decomposes the digital commons into its six foundational metrics: \textit{financial health} ($F_t$, $L_t$, and $H_t$) and \textit{social capital} ($P_t$, $C_t$, and $T_t$). To maximize data density and facilitate cross-metric comparison within a constrained vertical space, these metrics are visualized using horizon graphs.
A diverging red-to-blue palette encodes each metric on a shared \([-1,1]\) scale, enabling rapid identification of synchronized shocks or systemic decoupling across the dual-track commons.

Interactions within this view drive the analytical workflow through two complementary \mr{modes}. In the exploratory phase, brushing a time window updates the downstream \textit{Behavior View} and \textit{Communication Network}, allowing users to inspect historical agent interactions under empirical ground truth. To initiate a counterfactual backtest, users insert a customized intervention at a chosen timestamp, visually marked as a distinct red dot (Fig.~\ref{fig:interface}C2).
\mr{The simulation then produces alternative trajectories:}
the community-level timeline and horizon graphs update to overlay these counterfactual macro-metrics against historical baselines, while the \textit{Behavior View} and \textit{Communication Network} simultaneously transition to display the newly simulated meso- and micro-level data.

\subsection{Behavior View}

The \textit{Behavior View} supports meso-level behavioral analysis by tracking the evolving trajectories of persona groups and their heterogeneous responses to policy interventions (\textbf{T4}, \textbf{T5}). While existing approaches capture temporal shifts via asset-centric transfer flows~\cite{wen2023nftdisk} or aggregated Sankey diagrams~\cite{sheng2025eminds}, they inherently obscure the identity-preserving trajectories and multi-dimensional evolutions of specific cohorts by aggregating transitional volumes. To address this, we introduce a \textit{capsule-and-ribbon} design (Fig.~\ref{fig:interface}D).
By adapting rank-flow paradigms~\cite{tovanich2021miningvis}, this layout \mr{preserves cohort identity over time while revealing how persona trajectories bifurcate, enabling comparison of the three social-capital metrics and counterfactual branching.}

A customized ribbon chart plots persona groups along a chronological x-axis, vertically ranking them by overall social capital ($SC_t$). At each tick, a spatially efficient multi-attribute capsule glyph encodes the persona's state. Its total height represents absolute $SC_t$, while its interior is vertically partitioned into \textit{participation} ($P_t$), \textit{consensus} ($C_t$), and \textit{trust} ($T_t$) segments. This partitioning deliberately mirrors the visual hierarchy of the macro-level horizon graphs, ensuring cognitive consistency across views. Segment fill colors indicate metric polarity. To preserve identity and encode overall $SC_t$ polarity, the capsule's border applies the persona's palette using a solid stroke for positive $SC_t$ and a dashed stroke for negative $SC_t$.

Ribbons connect these capsules over time, with cross-overs illustrating rank transitions.
In the empirical ground-truth mode, the ribbon's fill defaults to the persona's palette to emphasize structural continuity (Fig.~\ref{fig:behaviorview}B1). Upon toggling to the counterfactual simulation mode, the ribbon dynamically shifts to encode the agent's simulated financial action at the originating node (Fig.~\ref{fig:behaviorview}B2). This design distinguishes data provenance while highlighting causal behavioral shifts.

Interaction techniques facilitate in-depth exploration. Selecting a persona icon on the left y-axis highlights its continuous temporal flow by attenuating others for focused tracking (Fig.~\ref{fig:behaviorview}B3). A dedicated toggle seamlessly switches between ground-truth and counterfactual modes, isolating the precise impact of injected governance rules.

\textit{Justification.}
To encode the three-dimensional metrics ($P_t$, $C_t$, $T_t$), we considered alternatives like heatmaps and treemaps. However, heatmaps compromise quantitative precision by over-relying on color channels, while treemaps disrupt cross-temporal tracking by dynamically altering spatial layouts. Our vertically partitioned capsule elegantly resolves these issues by maintaining a consistent spatial mapping and prioritizing geometric length over color intensity for accurate longitudinal comparison.

\begin{figure}[t]
\centering
\includegraphics[width=\columnwidth,
  alt={Communication Network visualization showing persona activity, interaction flows, temporal filters, and linked reasoning pathways.}]{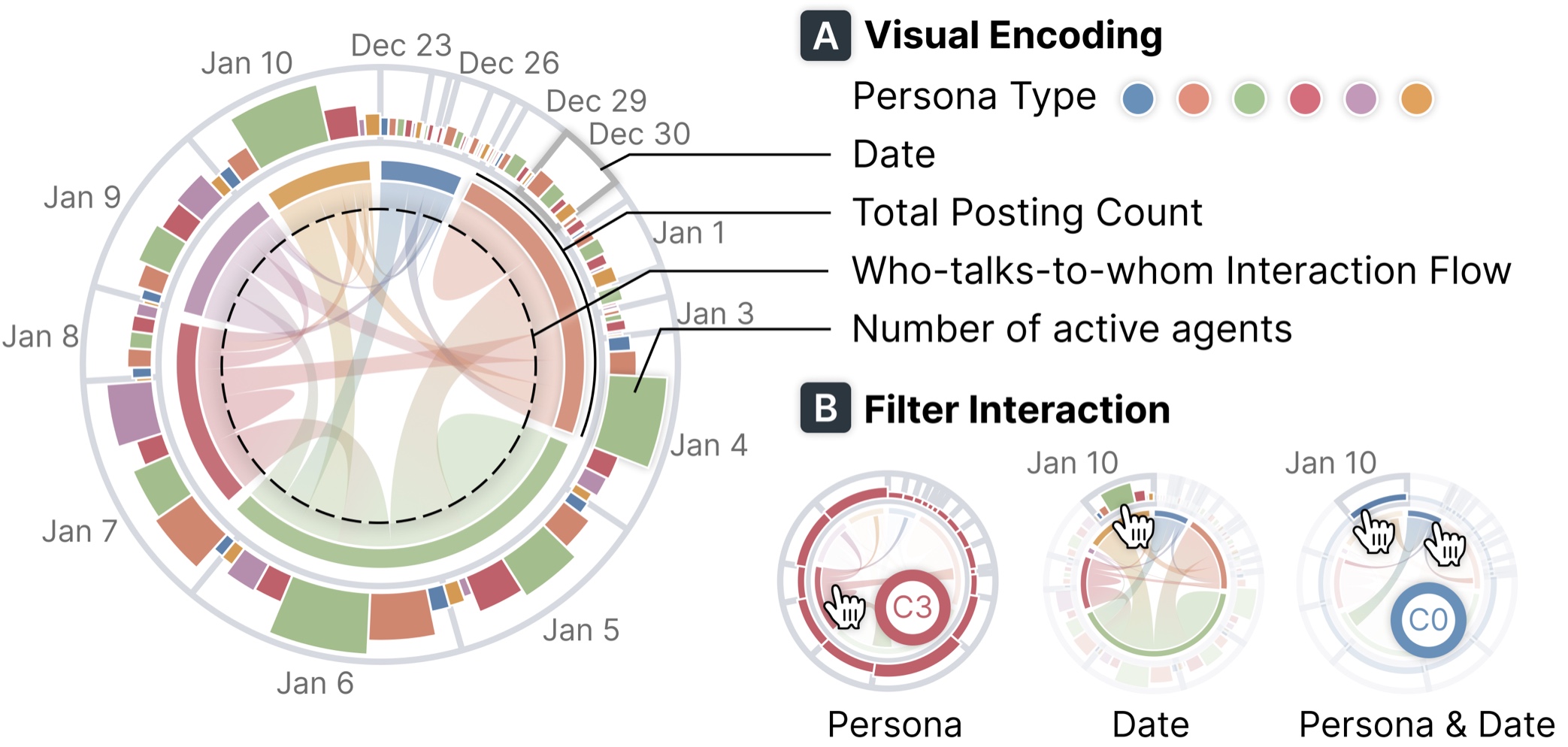}
\caption{Visual design of the \textit{Communication Network}. (A) Visual encodings show persona-driven social activation \mr{via} posting volumes, interaction flows, and active-agent counts. (B) Cross-view \mr{persona/date filters support focused} exploration of reasoning pathways.
}
\vspace{-2mm}
\label{fig:communi}
\end{figure}

\subsection{Communication Network}

The \textit{Communication Network} unpacks the micro-level communication dynamics and semantic rationales underlying meso-level behavioral shifts (\textbf{T5}, \textbf{T6}). This module juxtaposes aggregate interaction topologies with \mr{persona-grounded} reasoning pathways, \mr{enabling users to trace how policy interventions propagate through modeled agent behaviors to system-level metric shifts} (Fig.~\ref{fig:interface}E).

Inspired by~\cite{cao2023nfteller}, we designed a simplified multi-ring chord diagram to visualize the interaction topology (Fig.~\ref{fig:communi}A). Functioning primarily as a hierarchical information filter, its inner ring segments encode the total message volume per persona within the selected time window, with connecting ribbons mapping the communication flows; \mr{the outer ring, segmented by day as a circular bar chart, encodes each persona's daily active-agent count. The radial layout co-locates per-day activity with the persona interaction topology in one compact view, linking the outer and inner rings through shared persona colors rather than radial alignment, which favors space efficiency and cross-cohort comparison. The diagram controls the adjacent \textit{Reasoning Pathway} panel, which displays the agents' behavioral rationale and action distributions.}

Interactions facilitate a fluid drill-down from macro-topology to individual reasoning traces (Fig.~\ref{fig:communi}B). Intra-view, selecting an outer date arc or an inner persona segment filters the details panel to isolate rationales by day or cohort; combining both pinpoints a specific persona's cognition on an exact date. Cross-view coordination further drives the analytical workflow.
In counterfactual mode, injecting a policy event synchronizes updates across both the network and the reasoning pathways, allowing users to seamlessly trace how policy interventions catalyze micro-level cognitive shifts and dialogues (Fig.~\ref{fig:interface}E2).

\section{Evaluation}

We evaluated the effectiveness and usability of \textit{SocialFiVis} through two case studies, followed by a user study and semi-structured interviews.

\begin{figure*}[tb]
\centering
\includegraphics[width=\textwidth,
  alt={Case-study comparison of incentive-policy outcomes and persona-level deceptive financial behavior in SocialFiVis.}]{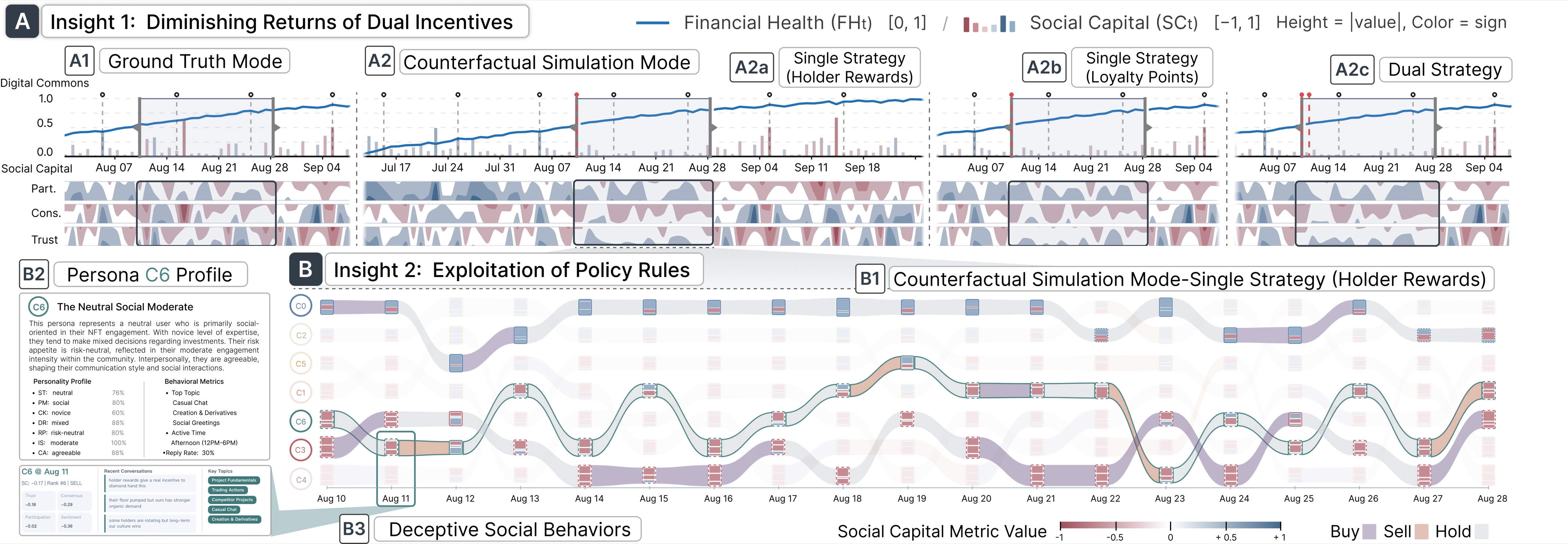}
\caption{Illustration of Case I. (A) Counterfactual simulations in the \textit{Event Timeline} reveal the diminishing marginal returns of the dual incentive strategy \mr{(A2c)} compared to single positive interventions \mr{(A2a, A2b)}. (B) Under a single incentive policy (B1), the \textit{Behavior View} exposes policy exploitation, illustrating how speculative agents (B2) leverage deceptive social behaviors (B3) for financial gain.}
\vspace{-2mm}
\label{fig:case1}
\end{figure*}

\subsection{Case Studies}
\label{sec:case}

\mr{$E_1$ and $E_5$} conducted exploratory case studies to evaluate \textit{SocialFiVis} in real-world analytical workflows.
\mr{As the simulated populations are instantiated from messaging members (Sec.~\ref{sec: persona extra}), the case studies characterize how social and financial dynamics co-evolve within this messaging population} under different governance interventions.

\subsubsection{Case I: Policy Efficacy and Diminishing Returns}
Focusing on policy strategy comparison during a stagnant market phase \mr{in the \textit{Mimic Shhans} community}, $E_1$ discovered that overlapping incentive strategies yield diminishing marginal returns and that positive macro-policies can be covertly exploited by deceptive individual behaviors (\textbf{DG2}, \textbf{DG3}).

\noindent\textbf{\mr{Actionable Insight 1: Stagger incentive releases rather than stacking them.}}
\mr{Examining the Aug. 10--28, 2022 phase, $E_1$ compared three counterfactual conditions: ``Holder Rewards'' alone, ``Loyalty Points'' alone, and the two combined. In the \textit{Event Timeline} and horizon graphs (Fig.~\ref{fig:case1}A), each incentive on its own lifted the social-capital trajectory, yet the combined run tracked the stronger single policy rather than exceeding it and left consensus slightly lower rather than stronger.}
$E_1$ noted, ``\textit{Deploying two favorable policies simultaneously violates the economic principle of marginal utility; it dilutes the intended signal and scatters community consensus.}'' The simulation \mr{accordingly guided $E_1$ toward staggering} incentive releases rather than stacking them.

\noindent\textbf{\mr{Actionable Insight 2: Monitor sentiment--action divergence during incentive campaigns.}}
Transitioning to the meso-level \textit{Behavior Ribbon} (Fig.~\ref{fig:case1}B), $E_1$ compared how different personas reacted to the injected liquidity of the ``Holder Rewards.'' While personas C0 and C3 stabilized the community as ``core promoters,'' $E_1$ observed anomalous fluctuations in C6, characterized by discordant color encodings between sentiment and financial action. By interactively filtering C6 in the \textit{Communication Network} and inspecting its micro-level \textit{Reasoning Pathway}, $E_1$ uncovered a highly deceptive pattern: C6 actively preached ``diamond hands'' (holding) in the chat to inflate sentiment while simultaneously executing aggressive sell orders.
$E_1$ concluded that \mr{operators should monitor sentiment--action divergence to flag such manipulative archetypes before they exploit macro-level policies.}


\subsubsection{Case II: \mr{Aggregate Resilience} and Latent Fragility}
$E_5$ \mr{examined how the \textit{Mfers} messaging cohort weathered} volatility from Dec. 23, 2022 to Jan. 10, 2023, and how macro-level momentum shaped social stability under adverse conditions (\textbf{DG1}, \textbf{DG3}).

\noindent\textbf{\mr{Actionable Insight 1: Watch trust--participation decoupling despite stable engagement.}}
Against a baseline of natural macro-level recovery (Founder's Return), $E_5$ evaluated resilience by comparing two counterfactual scenarios: a negative ``Key Member Exit'' shock and a positive ``Holder Rewards'' incentive. In the negative scenario, $E_5$ found that the \textit{participation} and \textit{consensus} metrics remained stable in the \textit{Event Timeline}, maintaining levels comparable to the positive intervention. In contrast, the \textit{trust} index exhibited a marked and isolated decline.
\mr{This structural decoupling shows $E_5$ that} strong macro-positive signals can sustain engagement and consensus even when interpersonal trust erodes under a localized negative shock.
\mr{$E_5$ accordingly flagged the isolated trust decline as a signal to monitor, since participation and consensus alone would mask it.}

\noindent\textbf{\mr{Actionable Insight 2: Prioritize vulnerable archetypes after a negative shock.}}
By cross-referencing the \textit{Behavior View} with the \textit{Reasoning Pathway}, $E_5$ \mr{found the trust erosion concentrated in specific archetypes rather than uniform.} ``Optimist'' believers (C3) held steady through the shock, \mr{whereas ``pessimist'' members (C0) reacted sharply, selling off and refuting peers as} their \textit{trust} \mr{plunged} and recovered only after the macro-milestone signal propagated (Fig.~\ref{fig:behaviorview}B3). \mr{``Band traders'' (C4) likewise turned net sellers, thinning support rather than cushioning the downturn. $E_5$ noted that this archetype-level attribution identifies C0 and C4 as the cohorts to monitor first when a negative shock lands.}

\subsection{User Study and Expert Interviews}

To evaluate the system across a broader audience, we conducted a user study supplemented by semi-structured interviews.

\subsubsection{Participants and Procedure}
In addition to $E_1$--$E_4$, we recruited \mr{two independent domain experts ($E_5$, $E_6$) and seven users ($U_1$--$U_7$).
By role, $E_5$, $U_1$, $U_4$, $U_6$, and $U_7$ were operations-experienced participants, whereas $E_6$, $U_2$, $U_3$, and $U_5$ were secondary users (analysts, traders, and ordinary members) who interpret or audit governance analyses.
This anchored the evaluation in our primary operator audience while probing broader applicability.}

The evaluation employed a \mr{two-track} procedure to accommodate participants' varying levels of prior engagement. For the 11 participants who did not conduct the exploratory case studies (i.e., $E_2$--$E_4$, $E_6$, and $U_1$--$U_7$), the session began with a 15-minute tutorial. They then completed 25 minutes of predefined analytical tasks using a think-aloud protocol. These tasks were directly mapped to our design goals: (1) correlating macro financial metrics with social capital (\textbf{DG1}); (2) using the sandbox to backtest historical interventions (\textbf{DG2}); and (3) tracing the heterogeneous decision pathways of specific personas (\textbf{DG3}).

$E_1$ and $E_5$ bypassed these predefined tasks, as their prior in-depth case studies (Sec.~\ref{sec:case}) far exceeded the complexity of the baseline tasks. Finally, all 13 participants completed a 5-point Likert scale questionnaire spanning five evaluation dimensions~\cite{lam2011empirical, hoffman2018metrics}, followed by a 10-minute post-study interview.

\subsubsection{Results and Feedback}
\begin{figure}[t]
\centering
\includegraphics[width=\columnwidth,
  alt={Questionnaire results for thirteen participants across twelve five-point Likert-scale questions.}]{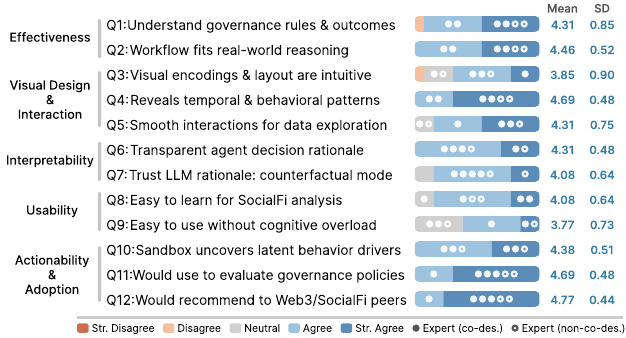}
\caption{User study results (N=13). Q1--Q12 are rated on a 5-point Likert scale, assessing \textit{SocialFiVis} across five dimensions.}
\vspace{-2mm}
\label{fig:userstudy}
\end{figure}

Quantitative results (see Fig.~\ref{fig:userstudy}) highlight the system's successes in \textit{visualization informativeness} ($M=4.69, SD=0.48$) and \textit{adoption willingness} ($M=4.69, SD=0.48$; $M=4.77, SD=0.44$ for \textit{recommendation}). Qualitative feedback supports these results. Participants consistently highlighted the system's ability to link macro market events with micro behavioral trajectories as its most valuable contribution. They found the \textit{Event Timeline} and \textit{Behavior View} particularly effective for this purpose. The \mr{two-phase agent} simulation was praised for capturing behavioral heterogeneity and decision-making nuance that purely statistical models cannot represent.
We synthesized the feedback below.

\noindent\textbf{System Usefulness and Effectiveness.} Participants highly rated the system's ability to model the interplay between governance rules and macro outcomes ($M=4.31, SD=0.85$). Notably, experts found the multi-level workflow closely aligned with their analytical reasoning processes ($M=4.46, SD=0.52$). $E_3$ and $E_4$ noted that it ``\textit{provides a quantitative baseline [for] backtesting,}'' while $E_2$ appreciated the persona classification, emphasizing that tracking who does what is ``\textit{genuinely helpful and intuitive for reporting to management.}''

\noindent\textbf{Visual Design and Interaction.} Participants praised the visualizations for their rich information density ($M=4.69, SD=0.48$). For instance, $U_1$ noted that the \textit{Behavior View} ``\textit{reveals social relationships... which is something you cannot see intuitively,}'' while $E_5$ found the \textit{Persona View}'s fine-grained classification highly aligned with real-world community experience. However, \mr{intuitiveness scored lower ($M=3.85, SD=0.90$), reflecting a trade-off with informativeness: the dense, multi-view layouts mobilize cognitive resources but carry a learning curve, particularly for community builders and other less data-oriented users who focus on engagement rather than analysis. This adaptation is expected, as their prior practice often leaned on trial-and-error, intuition, and one-on-one interviews rather than systematic visual analytics.} $U_4$ captured the balance: the \textit{Event Timeline} ``\textit{gives you the big picture,}'' though its density demands familiarization.

\noindent\textbf{Interpretability and Trust.} The system successfully established transparency in agent decision-making ($M=4.31, SD=0.48$) and trust in the \mr{agent simulation} ($M=4.08, SD=0.64$). Observing a simulated agent's actions, $E_5$ remarked, ``\textit{The AI is remarkably true-to-life... its stated sentiment, reasoning, and final action all line up.}'' To further enhance this utility, $E_1$ suggested incorporating a proactive policy recommendation feature. Since formulating governance policies is a high-frequency daily task, such automated guidance could meaningfully streamline operators' routine analysis workflows.

\noindent\textbf{Usability.} Learnability scored positively ($M=4.08, SD=0.64$), though cognitive load ratings ($M=3.77, SD=0.73$) reflected the aforementioned learning curve. Crucially, usability perceptions varied by professional role. For \mr{data-savvy operators and analysts}, the density is a strategic asset; $E_3$ found it ``\textit{genuinely useful for my day-to-day work,}'' and $E_2$ noted that ``\textit{for someone in a marketing role, this is very usable,}'' even expressing a willingness to pay for a commercial version.
\mr{Conversely, less data-oriented users would benefit from a simplified dashboard to ease their initial cognitive load.}

\noindent\textbf{Actionability and Adoption.} The sandbox's ability to uncover latent catalysts scored highly ($M=4.38, SD=0.51$). $E_6$ expressed strong intent to recommend the system to executive leadership, emphasizing that its rich insights make it highly ``\textit{suited for decision-makers}.'' However, participants also identified data dependency as a critical prerequisite. $E_5$ and $U_7$ cautioned that during data collection, operators must carefully filter out chatbots and airdrop hunters (speculators) to prevent them from polluting the user personas and skewing the simulation outcomes.

\section{Discussion}
We reflect on the broader implications of \textit{SocialFiVis}, detailing its methodological generalizability and limitations guiding future research.

\subsection{Significance and Generalizability}
Beyond addressing specific governance challenges in tokenized communities, the core contribution of \textit{SocialFiVis} lies in its highly transferable methodology for analyzing the co-evolution of social and financial dynamics in broader decentralized environments.

\noindent\textbf{Significance of the Work.} \textit{SocialFiVis} operationalizes the IAD framework into a visual analytics pipeline, bridging the micro-macro gap by coupling \textit{social capital} with \textit{financial health}. Crucially, the counterfactual sandbox provides a risk-free environment for policy testing.
This directly addresses a longstanding practitioner pain point: the ``attribution problem'' of tracing mechanistic pathways between specific community interventions and subsequent financial outcomes.
Furthermore, \mr{by grounding agents in LLM-derived, expert-validated personas, the system represents behavioral heterogeneity within the retained messaging cohort and supports inspection of modeled decision rationales.}

\noindent\textbf{Generalizability of the Methodology.} We designed the system's core algorithms to generalize far beyond financialized ecosystems. Because the IAD framework inherently governs common-pool resources~\cite{ostrom2009institutional}, our approach is naturally applicable to various decentralized organizations, creator economies, and open-source communities. Specifically, the persona extraction and PRA pipeline can serve as a template to simulate heterogeneous populations in corporate organizational behavior or consumer research. Additionally, our soft-penalty geometric blending model can be adapted to quantify other forms of intangible capital~\cite{hunter2005measuring}, such as user engagement in online learning platforms or the health metrics of corporate cultures.

\subsection{Limitations and Future Work}
While our evaluation confirms the system's utility, it also highlights limitations that pave the way for future enhancements.

\noindent\textbf{Simulation Boundaries and \mr{Predictive Validity}.}
\mr{Instantiated from messaging members, the simulation excludes silent and near-silent accounts below our persona-inference threshold (Sec.~\ref{sec: persona extra}); it therefore reflects expressed dynamics, not silent disengagement~\cite{soroka2006invisible, nielsen2006participation}.
This exclusion bounds the aggregate resilience reported in our case studies (Sec.~\ref{sec:case}).
Within this cohort, the PRA pipeline produces agents' posts and trades from a reasoning step but maintains no \textit{persistent belief state} separable from expression. Such a state would model agents who hold views without posting and make discrepancies between words and actions a systematic, auditable signal for users rather than an incidental finding.}
\mr{This scope also shapes validation. Because the simulated interventions never occurred, we follow standard ABM practice and validate aggregate trends against ground truth (Table~\ref{tab:validation}) rather than claiming persona-level fidelity to real users. We therefore present outcomes as exploratory reasoning aids instead of forecasts~\cite{fagiolo2007critical, windrum2007empirical}.
Even so, communicating aggregate uncertainty to calibrate user trust remains an open challenge~\cite{wall2024trust}.
The LLM may also have seen these communities' public messages in training, a general concern for pretrained models; we temper but cannot fully remove this exposure by confining the LLM to a study-specific coding scheme, while the mechanism-guided runtime generates the simulated behaviors and trajectories.}

\noindent\textbf{System Complexity and Learning Curve.}
The system's high information density entails a learning curve. \mr{Although its coordinated views are organized around progressive disclosure~\cite{shneiderman1996eyes, elmqvist2010hierarchical}, the lower intuitiveness score ($M=3.85$) shows that the entry state remains demanding, especially for less data-oriented users.
A deployment-ready design therefore requires a compact summary layer (e.g., directional bias indicators, aggregate sentiment--action divergence, and flagged milestones) before users opt into the denser persona, behavior, and reasoning views.
The same practical lens applies to evaluation: although we recruited independent operators and secondary users, the co-designers' involvement may bias feedback favorably, and a larger participant pool studied in operators' day-to-day workflows would further corroborate the findings.}

\section{Conclusion}
In this paper, we presented \textit{SocialFiVis}, a visual analytics sandbox that operationalizes the IAD framework to quantify the dual-track digital commons and \mr{model} the non-linear interplay between governance rules, individual behaviors, and economic outcomes.
\mr{Its two-phase simulation pairs LLM-derived, expert-validated personas with a mechanism-guided PRA runtime, letting users trace how interventions propagate through agent rationales into shifts within the messaging cohort.
Across two case studies and expert interviews, \textit{SocialFiVis} surfaced actionable insights, turning theoretical governance into concrete behavioral attribution.}
Ultimately, it establishes a novel foundation for simulation-based decision support at the intersection of visual analytics, agent-based modeling, and digital commons governance.




\section*{Ethics Statement}
Our formative and user studies were approved by HKUST's Human and Artefacts Research Ethics Committee under one protocol; all participants gave voluntary informed consent and are reported anonymously.

\section*{Supplemental Materials}
Supplemental materials, including appendices, task traceability, metric formulation, sensitivity analyses, persona and simulation validation, user-study instruments, and a demo video, are available via \href{https://doi.org/10.17605/OSF.IO/XYWMZ}{OSF}.

\acknowledgments{%
  The authors thank Lue Shen for valuable suggestions on the seven-dimensional persona codebook, and Xiyuan Wang for assistance with several of the illustrative elements in our figures.%
}

\bibliographystyle{abbrv-doi-hyperref}
\bibliography{Reference}

\appendix
\crefalias{section}{appendix}

\section{Appendix}
\renewcommand{\arraystretch}{1.05}

\begin{figure}[bh]
\centering
\includegraphics[width=\columnwidth,
  alt={Bar chart summarizing expert validation of the seven-dimensional persona codebook.}]{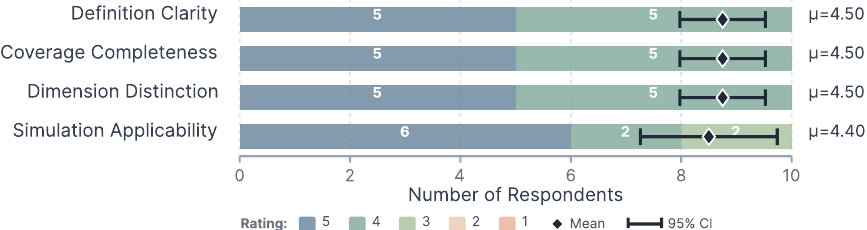}
\caption{Expert validation of the 7D persona codebook (N=10).}
\label{fig:expert_validation}
\end{figure}

\begin{table}[htbp]
\caption{Statistics of the data filtering and routing pipeline.}
\label{tab:filtering_stats}
\centering
\footnotesize
\resizebox{\columnwidth}{!}{%
\begin{tabular}{@{}lrrrrrrr@{}}
\toprule
& \multicolumn{2}{c}{\textbf{Raw Dataset}} & \multicolumn{2}{c}{\textbf{Messaging Filtering}} & \multicolumn{3}{c}{\textbf{Routing (Active Nodes)}} \\
\cmidrule(lr){2-3}\cmidrule(lr){4-5}\cmidrule(l){6-8}
\multicolumn{1}{c}{\multirow{2}{*}{\textbf{Community}}}
& \multicolumn{1}{c}{\multirow{2}{*}{\shortstack{\textbf{Text}\\\textbf{Msgs}}}}
& \multicolumn{1}{c}{\multirow{2}{*}{\textbf{Nodes}}}
& \multicolumn{1}{c}{\textbf{Dropped}}
& \multicolumn{1}{c}{\textbf{Retained}}
& \multicolumn{1}{c}{\textbf{High-Conf.}}
& \multicolumn{1}{c}{\textbf{Low-Conf.}}
& \multicolumn{1}{c}{\textbf{Manual}} \\
& & &
\multicolumn{1}{c}{($<15$)}
& \multicolumn{1}{c}{($\geq15$)}
& \multicolumn{1}{c}{($\geq0.60$)}
& \multicolumn{1}{c}{($<0.60$)}
& \multicolumn{1}{c}{(Subset)} \\
\midrule
\textbf{Mfers} & 47,646 & 395 & 194 (49.1\%) & 201 (50.9\%) & 123 (61.2\%) & 78 (38.8\%) & 24 (11.9\%) \\
\textbf{Mimic Shhans} & 41,740 & 435 & 218 (50.1\%) & 217 (49.9\%) & 114 (52.5\%) & 103 (47.5\%) & 33 (15.2\%) \\
\bottomrule
\end{tabular}%
}
\end{table}

\end{document}